----------------------------------------------------------------

 \documentstyle[12pt]{article}
  \newcommand{\be}{\begin{equation}}
  \newcommand{\ee}{\end{equation}}
  \newcommand{\ba}{\begin{array}{llll}}
  \newcommand{\ea}{\end{array}}
  \renewcommand{\thefootnote}{\fnsymbol{footnote}}
\begin{document}
 \topmargin 0pt
 \oddsidemargin=-0.4truecm
 \evensidemargin=0.4truecm
 \newpage
 \setcounter{page}{0}
 \begin{titlepage}
 \begin{flushright}
 LMU-08/92 \\
  July 1992
 \end{flushright}
 \vspace{1.0cm}
 \begin{center}
  {\Large\bf Radiative Origin of the Fermion Mass Hierarchy:\\
  \vspace{3mm}
 A Realistic and Predictive Approach} \footnote{Talk given by Z.Berezhiani
 at XV International Warsaw Meeting "Quest for Links to New
 Physics", Kazimierz, Poland, May 25-29, 1992.}\\ [6mm]
 {\large  Zurab Berezhiani} \footnote{Alexander
 von Humboldt fellow. }\footnote{
 E-mail: zurab@hep.physik.uni-muenchen.de, vaxfe::berezhiani}\\ [2mm]
 {\em Sektion Physik, Universit\"at M\"unchen,
 D-8000 M\"unchen 2, Germany\\
 Institute of Physics, Georgian Academy
 of Sciences, SU-380077 Tbilisi, Georgia}\\ [5mm]
 and\\ [5mm]
 {\large Riccardo Rattazzi} \\ [2mm]
 {\em Theoretical Physics Group, Lawrence Berkeley Laboratory, Berkeley,
 CA 94720, USA} \\
  \end{center}
 \vspace{2mm}
 \begin{abstract}
  The up-down splitting within quark families
 increases with the family number: $m_u \sim m_d,~ m_c > m_s,~ m_t
 \gg m_b,$. We show an approach that realizes this feature of the
 spectrum in a natural way.
 We suggest that the mass hierarchy is first generated by radiative effects
 in a sector of heavy isosinglet fermions, and then
 projected to the ordinary light fermions by means of a seesaw mixing.
 The hierarchy appears then {\em inverted} in the light fermion sector.
 We present a simple left-right symmetric gauge model in which
 the $P$- and $CP$-parities and an isotopical "up-down"
 symmetry are softly (or spontaneously) broken in the Higgs
 potential. Experimentally consistent predictions are obtained.
 The Cabibbo angle is automatically in the needed
 range: $\Theta_C \sim 0.2$. The top quark is naturally heavy, but
 not too heavy: $m_t < 150$ GeV.
 \end{abstract}
 \end{titlepage}
 \newpage
 \baselineskip0.7cm
 \renewcommand{\thefootnote}{\arabic{footnote})}
 \setcounter{footnote}{0}

 Although the idea of radiatively generated
 fermion mass hierarchy is very attractive, it is difficult to implement it
 in a realistic way. For instance, it is generally problematic to understand
 the experimental value of the Cabibbo angle and the large {\em top-bottom}
 splitting. In addition dangerous FCNC's have to be kept under control.
 Recently$^1$, however, a new approach to the fermion mass puzzle
 has been suggested. In this approach
 the mass hierarchy is first radiatively generated in a hidden sector of
 hypothetical heavy fermions and then transferred to the visible quarks
 and leptons by means of a {\em universal seesaw} mechanism$^2$. Providing a
 qualitatively correct picture of quark masses and mixing, this approach
 solves many problems of the previous models$^{3,4}$ of radiative
 mass generation. In particular, the correct value of the Cabibbo
 angle can be accommodated, without trouble for the perturbative expansion.
 Moreover, within the seesaw approach, the effective low energy theory,
 after integrating out the heavy fermions, is simply the standard model
 with one Higgs doublet (and with {\em definite} Yukawa couplings).
 Thus, flavour changing
 phenomena, typical of the direct models$^4$ of radiative mass
 generation, are naturally suppressed.

  The key idea of the model$^1$ is to suppose the existence of weak
 isosinglet heavy fermions (Q-fermions) in one-to-one correspondence with
 the light ones. The model$^1$ has a field content such that only one
 family (namely the first) of Q-fermions becomes massive at the tree
 level.  The 2$^{nd}$ Q-family gets a mass
 at the 1-loop level and the 3$^{rd}$ only at 2 loops.
 Because of the seesaw mechanism$^2$,  the mass of any usual quark
 or lepton is inversely proportional to the mass of its heavy
 partner. Thus the mass hierarchy between the families of light fermions
 is {\em inverted} with respect to the hierarchy of Q-fermion families. This
 feature is very attractive for the following reason.
 Experimentally  we observe a
 small mass   splitting within the lightest quark family ($u$ and $d$) and
 an increasing splitting from family to family, with the up-quark masses
 growing faster: $1 \sim m_u/m_d< m_c/m_s< m_t/m_b$.  In our approach
 it is natural to have $m_u\sim m_d$, since these masses are determined by
 the tree level masses of the heaviest Q-fermions. On the other hand,
 the increasing splitting can be related to the difference
 between the loop-expansion parameters in the up and down quark sectors.

 In what follows, we show that the simplest and most economical
 version of the model$^1$ provides a predictive ansatz for the quark
 mass matrices. We assume that the ``isotopical'' discrete symmetry $I_{UD}$
 between up and down quark sectors, as well as the left-right symmetry
 $P_{LR}$ and $CP$-invariance, is violated {\em only} in the loop expansion,
 due to soft (or spontaneous) breaking in the  Higgs potential. The appearance
 of {\em both} the mass splitting within the lightest family ($m_d/m_u=1.5-2$)
 and the large (compared to the other mixing angles) value of the Cabibbo angle
 ($\sin\Theta_C\simeq 0.22$) is determined by
 the properties of the seesaw ``projection''. The troubles for the
 perturbation  expansion are then avoided.
 The model leads to some successful predictions
 for the quark mass and mixing pattern. We shall discuss them below.

  Let us consider the simple left-right symmetric model based on the gauge
 group
  $G_{LR}=SU(2)_L \otimes SU(2)_R \otimes U(1)_L \otimes U(1)_R \otimes U(1)_
 {\bar{B}-\bar{L}}$,
 suggested in$^1$. The left- and right-handed components of  usual quarks
 $q_i=(u_i,d_i)$  and their heavy partners $ Q_i=U_i,D_i $ are taken
 in the following representations:
  \be
  \ba
  q_{Li}(I_L=1/2,~\bar{B}-\bar{L}=1/3),~~~~~q_{Ri}(I_R=1/2,
  ~\bar{B}-\bar{L}=1/3) \\
  U_{Li}(Y_L=1,~~~\bar{B}-\bar{L}=1/3),~~~~~U_{Ri}(Y_R=1,~~~
  \bar{B}-\bar{L}=1/3) \\
  D_{Li}(Y_L=-1,~\bar{B}-\bar{L}=1/3),~~~~D_{Ri}(Y_R=-1,
  ~\bar{B}-\bar{L}=1/3) \\
  \ea
  \ee
  where i=1,2,3 is the family index (the indices of colour $SU(3)_c$ are
  omitted).
 Only the nonzero quantum numbers are shown in the brackets:
 $I_{L,R}$ are the $SU(2)_{L,R}$ weak isospins and $Y_{L,R}$ are
 the $U(1)_{L,R}$ hypercharges. Let us also introduce {\em one} additional
 family of fermions with $\bar{B}-\bar{L}=1/3$ and with the following
  hypercharges:
  \be
  \ba
  p_L(Y_L=-1/2,~Y_R=3/2),~~~~~
  p_R(Y_L=3/2,~Y_R=-1/2) \\
  n_L(Y_L=1/2,~Y_R=-3/2),~~~~~
  n_R(Y_L=-3/2,~Y_R=1/2)
  \ea
  \ee
  The scalar sector of the theory consists of
  \be
  \ba
  H_L(I_L=1/2,~~Y_R=1),~~~~~~~~~~~~~~~~~H_R(I_R=1/2,~~~Y_L=1) \\
  T_{uL}(Y_L=-2,~~\bar{B}-\bar{L}=-2/3),~~~~~~T_{uR}(Y_R=-2,
  ~~\bar{B}-\bar{L}=-2/3) \\
  T_{dL}(Y_L=2,~~~\bar{B}-\bar{L}=-2/3),~~~~~~~~
  T_{dR}(Y_R=2,~~~\bar{B}-\bar{L}=-2/3) \\
  \Phi(Y_L=2 ,Y_R=-2),~~~~~~~~~~~~~~~~~~~~~~\varphi(Y_L=1/2,Y_R=-1/2),~ \\
  \Omega(Y_L,Y_R=1/2,~\bar{B}-\bar{L}=-1)
  \ea
  \ee
  where the T-scalars are supposed to be colour triplets. Let us impose also
  $CP, P_{LR}$ and $I_{UD}$ discrete symmetries. $P_{LR}$$^6$,
   essentially parity, and $CP$ act in the usual way. The
  isotopical ``up-down'' symmetry $I_{UD}$ is defined by
  \be
  \ba
  U_{L,R}\leftrightarrow D_{L,R},~~~p_{L,R}\leftrightarrow n_{L,R},~~~
  H_{L,R}\leftrightarrow \tilde{H}_{L,R}=i\tau_2 H_{L,R}^*, \\
  T_{L,R}^u\leftrightarrow T_{L,R}^d,~~~\Phi\leftrightarrow \Phi^*,~~~
  \varphi\leftrightarrow\varphi^*,~~~
 A_{L,R}^{\mu}\leftrightarrow -A_{L,R}^{\mu}
  \ea
  \ee
  where $A_{L,R}^{\mu}$ are the gauge bosons of $U(1)_{L,R}$. Then the most
  general Yukawa couplings consistent with gauge invariance,
 $I_{UD}, P_{LR}$ and $CP$ are
 \be
 \ba
 {\cal L}_1=\Gamma_{ij}(\bar{q}_{Li}\,U_{Rj}\,\tilde{H}_L+
 \bar{q}_{Li}\,D_{Rj}\,H_L)+(L\leftrightarrow R)+h.c.   \\
 {\cal L}_2=\lambda_{ij}(U_{Li}CU_{Lj}T_{uL}+D_{Li}CD_{Lj}T_{dL})+
 (L\leftrightarrow R)+h.c. \\
 {\cal L}_3=h(\bar{p}_Lp_R\Phi^*+ \bar{n}_Ln_R\Phi)+h_i(\bar{U}_{Li}p_R
 \varphi^*+
 \bar{D}_{Li}n_R\varphi)+(L\leftrightarrow R)+h.c.
 \ea
 \ee
 where C is the charge conjugation matrix. The coupling constants $h, h_i,
 \lambda_{ij}, \Gamma_{ij} (i,j=1,2,3)$ are {\em real} due to CP-invariance
 ($\lambda_{ij}=-\lambda_{ji}$, since the
 T-scalars are colour triplets). In
 what follows we do not make any particular assumption on their structure.
 We only suppose that they are {\em typically} $O(1)$, just like the gauge
 coupling constants. Without loss of generality, by a suitable redefinition of
 the fermion basis, we can always take $h_2,h_3=0$, $\lambda_{13}=0$,
 $\Gamma_{12},\Gamma_{13},\Gamma_{23}=0$. In what follows we use this basis.

  Let us suppose that the discrete symmetries $CP, P_{LR}$ and $I_{UD}$
  are {\em softly} broken only by the bilinear and trilinear terms in the Higgs
  potential
  \footnote{Actually, this symmetries could be spontaneously broken at
 the price of introducing $P_{LR}-$ and $I_{UD}-$odd real
 scalars$^1$. The consequences, as far as fermion masses
 are concerned, would be unchanged.}.
  These are given by
  \be
  {\cal V}_3=\Lambda_u T_{uL}^* T_{uR}\Phi+\Lambda_d T_{dL}^*T_{dR}\Phi^* +h.c.
  \ee
 where the coupling constants $\Lambda_{u,d}$ are generally {\em complex},
 violating thereby both $CP$ and $P_{LR}$ invariances.

  The VEVs $\langle\Phi\rangle=v_{\Phi}$ and $\langle\varphi\rangle=
  v_{\varphi},~ v_{\Phi}\gg v_{\varphi},$~ break $U(1)_L\otimes U(1)_R$
 down to
  $U(1)_{L+R}$ (the VEV of $\Omega$ then breaks $U(1)_{L+R}\otimes
  U(1)_{\bar{B}-\bar{L}}$ to the usual $U(1)_{B-L}:B-L=Y_L+Y_R+\bar{B}-
 \bar{L}$ ).
  The fermions $p$ and $n$ become massive, $M_p=M_n=hv_{\Phi}$, and the
  Q-fermions of the first family, $U_1$ and $D_1$ get
  masses $M\cong h_1^2 v_{\varphi}^2/hv_{\Phi}$
  due to their {\em seesaw} mixing with the former ones (interactions
 ${\cal L}_3$ in
 (5)). At the same time the coloured scalars $T_{uL}-T_{uR}$ and
 $T_{dL}-T_{dR}$ get mixed due to the interaction terms
 in (6). At this point,
 radiative mass generation proceeds, following the chain
 $U_1\rightarrow U_2\rightarrow U_3,~~~D_1\rightarrow D_2 \rightarrow D_3$ .
  The Q-fermion mass matrices generated from the loop corrections due to
 ${\cal L}_2$ in (5) can be presented in the following form:
  \be
  M_{U,D}=M(\hat{P}_1+e^{-i\omega_{u,d}}\,\xi_{u,d}\,\tilde{\lambda}\,
 \hat{P}_1\,
  \lambda+C_{u,d}\xi_{u,d}^2\,\tilde{\lambda}^2 \,
 \hat{P}_1\,
  \lambda^2+\cdots)
  \ee
  where $\hat{P}_1=diag(1,0,0)$ is a 1-dimensional projector and
 $\omega_{u,d}=-\arg \Lambda_{u,d}$. The loop expansion factors are
  \be
  \xi_{q}=\frac{1}{8\pi^2}\sin 2\alpha_{q}\,\log R_{q},
  ~~~~~~R_{q}=(M_{+}^{q}/M_{-}^{q})^2
  \ee
  where $M_{+}^{q}, M_{-}^{q}$ are the eigenvalues of the mass matrices of the
 scalars $T_{qL}-T_{qR}$, $q=u,d$, and $\alpha_{q}$ are the corresponding
 mixing angles. In a ``reasonable'' range of parameters ($1<R<10$) the 2-loop
 factor $C(R)=C(1/R)$ is practically constant$^4$: $C_{u,d}\simeq 0.65$.
 Eq.(8) is valid in the natural regime $M<M_{+}^{q},M_{-}^{q}<M_{p}$.

    Apart from small $O(\xi_{u,d}^2)$ 1-3 entries, the
 matrices $M_{U,D}$ are {\em diagonal}. Then the mass hierarchy between
 the three families of Q-fermions is given by
 $1:x^{-1}\varepsilon_{u,d}:\varepsilon_{u,d}^2$,
  where we defined $x=\sqrt{C}\lambda_{23}/\lambda_{12}$ and
 $\varepsilon_{u,d}=\sqrt{C}\lambda_{12}\lambda_{23}\xi_{u,d}\sim 10^{-2}-
 10^{-1}$. The parameters $\varepsilon_{u}$ and $\varepsilon_{d}$
 are the effective loop-expansion parameters, respectively for the
 up and down sectors.

  The VEVs $\langle H_L\rangle=(0,v_L)$ and $\langle H_R\rangle=(0,v_R),~~~
  v_R\gg v_L=(2\sqrt{2}\,G_F)^{-1/2}\approx 175$ GeV, break the intermediate
  $SU(2)_L\otimes SU(2)_R\otimes U(1)_{B-L}$ symmetry down to $U(1)_{em}$.
 Then the ordinary quarks $q=u,d$ acquire masses due to their seesaw mixing
 with the heavy fermions $Q=U,D$ (interactions ${\cal L}_1$ in eq.(5)). The
  whole mass matrix for up-type quarks, written in block form, is
  \be
  (\bar{u},\bar{U})_L\left(
  \begin{array}{cc}
  0 & \Gamma v_L \\
  \tilde\Gamma v_R & M_U
  \end{array}\right)
  \left(
  \begin{array}{c}
  u \\ U
  \end{array}\right)_R
  \ee
 and analogously for down-type quarks.
  When $M_{U,D}\gg v_R,v_L$, the resulting mass matrix for the light states
 is given by the seesaw formula
 \be
 M_{light}^{u,d}=v_L v_R \Gamma M_{U,D}^{-1}\tilde\Gamma
 \ee
 Substituting here eq.(7) we find, in the explicit form,
 \be
  M_{light}=\frac{m}{\varepsilon^{2}}\left(\begin{array}{ccc}
 \varepsilon^2\gamma_{11}^{2}&
 \varepsilon^2\gamma_{11}\gamma_{21}&
 \varepsilon^2\gamma_{11}\bar\gamma_{31}\\
 \varepsilon^2\gamma_{11}\gamma_{21}&
 \varepsilon xe^{i\omega}\gamma_{22}^{2}+\varepsilon^2\gamma_{21}^{2}&
 \varepsilon xe^{i\omega}\gamma_{22}\gamma_{32}+
 \varepsilon^2\gamma_{21}\bar\gamma_{31}\\
 \varepsilon^2\gamma_{11}\bar\gamma_{31}&
 \varepsilon xe^{i\omega}\gamma_{22}\gamma_{32}+
 \varepsilon^2\gamma_{21}\bar\gamma_{31}&
 1+\varepsilon xe^{i\omega}\gamma_{32}^{2}+\varepsilon^2\bar\gamma_{31}^{2}\\
 \end{array}\right)
 \ee
 where $m=\Gamma_{33}^{2}v_{L}v_{R}M^{-1}$,
  $\gamma_{ij}=\Gamma_{ij}/\Gamma_{33}$ and
  $\bar\gamma_{31}=\gamma_{31}+\sqrt{C} x^{-1}$;
 $\varepsilon=\varepsilon_{u,d}$, $\omega=\omega_{u,d}$ for the up and down
 quarks, respectively.

  It is obvious, from the measured values of quark masses, and
  from (11), that $\varepsilon_{u}\ll\varepsilon_{d}\ll 1$.
 The up quark mass matrix $M_{light}^{u}$ is almost
 diagonal. Neglecting $\sim\varepsilon_{u}$ corrections we have
 $m_{u}=m\gamma_{11}^{2}$, $m_{c}=xm\gamma_{22}^{2}\varepsilon_{u}^{-1}$
 and $m_{t}=m\varepsilon_{u}^{-2}$. Thereby, the quark mixing pattern
 is determined {\em essentially} by the down quark mass matrix $M_{light}^{d}$,
 where $m_{b}\approx m\varepsilon_{d}^{-2}$. The contributions to the
 parameters of the CKM matrix from $M_{light}^{u}$ are typically suppressed
 by the factor $\varepsilon_{u}/\varepsilon_{d}$ and we neglect them.
 After some algebra one can obtain:
 \be
 V_{us}\approx\sqrt{\frac{m_d}{m_s}\mid 1-\frac{m_u}{m_d}e^
 {i\delta}\mid}
 \ee
 \be
 V_{ub}\approx\frac{\bar\gamma_{31}}{\gamma_{11}}\frac{m_u}{m_b},~~~~~~
 V_{cb}\approx\frac{m_d}{m_u}\left(\sqrt{\frac{m_s}{m_d}}V_{ub}+
 \frac{\gamma_{32}}{\gamma_{22}}\frac{m_s}{m_b}e^{i\omega_d}\right)
 \ee
 where $\delta = -\omega_{d}+\arg (xe^{i\omega_{d}}\gamma_{22}^{2}+
 \varepsilon_{d}\gamma_{21}^{2}) \approx -\omega_{d}+
 \arg (1+e^{i\omega_{d}})$ is a phase
 measuring the amount of $CP$-violation in the CKM matrix. Within
 uncertain (but
 supposed to be $\sim 1$) numerical factors the formulae (13) fit the
 experimental values of $V_{ub}$ and $V_{cb}$ (notice that for $\Gamma_{32}=0$
 one has $V_{ub}/V_{cb}=m_u/\sqrt{m_dm_s}=0.11-0.15$). The small values of
 $V_{ub}$ and $V_{cb}$
 imply that the corresponding entries in $M_{light}^{d}$ {\em cannot}
 significantly affect the eigenvalues. As for the
 1-2 mixing, the situation is different. The relation $m_u\not = m_d$ requires
 a correction to $m_d$ from the 1-2 entry in $M_{light}^d$. This correction
 is of the right order of magnitude, provided
 $\Gamma_{21}/\Gamma_{11}=O(\sqrt{m_s/m_u})\approx 6$. We consider such a
 spread, in the value of the Yukawa coupling,  perfectly acceptable.
 As a result, the formula (12) appears which implies the
 Cabibbo angle to be in the needed range: $V_{us}=0.22\pm 0.07$ within all
 uncertainties.
 The comparison of (12) with the experimental value $V_{us}\approx 0.22$
implies
 a large $CP$-phase, $\delta\sim 1$, in agreement with the recent data.

 From the mass matrices (11) one can also derive the relations
 \be
 \frac{\varepsilon_d}{\varepsilon_u}=\frac{m_um_c}{m_dm_s}=
 \sqrt{\frac{m_t}{m_b}}
 \ee
 which allows to calculate the top quark mass using the known masses$^7$ of
 the other quarks. The large value of the former implies that
 the ``seesaw'' corrections$^8$ to equation (10) have to be taken into
 account. Doing so, we obtain the physical top quark mass
 \be
 m_{t}^*=m_{t}^{0}\left[1+
 \left(\frac{m_{t}^{0}}{\Gamma_{33}v_L}\right)^{2}\right]^{-1/2}
 \ee
 where $m_{t}^{0}$ is the ``would be'' mass, calculated from eq.(14).
 Obviously, the analogous corrections are negligible for other quark
 masses since we demand all $\Gamma$'s to be $\sim 1$. On the other
 hand, from (11) one can easily derive that $\Gamma_{21}/\Gamma_{33}\approx
 \varepsilon_{d}^{-1}\sqrt{m_dm_s/m_um_b}\geq 0.17\varepsilon_{d}^{-1}$.
 In order to be consistent with perturbation theory
 we assume
 that all the Yukawa coupling constants, including $\Gamma_{21}$ and
 $\lambda$'s, are less than 2. This implies $\Gamma_{33}\leq 1$.
 Consequently,
 from (14) and (15) we obtain $m_t^*=50-150$ GeV.
 The large spread here is related mainly with the uncertainties in the
 light quark masses.
 It is also interesting to turn the logic around
 and say that the experimental lower bound$^9$  $m_t^*>91$ GeV favours
 the lower values of $m_d/m_u$ and $m_s$ among those allowed in$^7$.

    The inclusion of leptons in this model is straightforward
 and will be presented elsewhere.
 In fact  $U(1)_{\bar{B}-\bar{L}}$ can be unified with $SU(3)_c$
 within Pati-Salam$^{10}$ type SU(4). The $U(1)_L\otimes U(1)_R\otimes I_{UD}$
 part can also be enlarged to $SU(2)_{L}^{\prime}\otimes SU(2)_{R}^{\prime}$,
 in which case the isotopical symmetry is obviously continuous.

    Last but not least we wish to remark that in our approach the
 strong $CP$-problem can be automatically solved {\em without} axion. Owing to
 $P$ and/or $CP$-invariances the initial $\Theta_{QCD}=0$ and
 $\Theta_{QFD}=\arg Det\hat{M}$, where $\hat M$ is the {\em whole} mass matrix
 $\hat{M}$ of all fermions $q$, $Q$ and $p,n$ is also
 vanishing at tree level, because of the seesaw pattern$^{11}$. The loop
 corrections
 can provide, however, $\bar{\Theta}=10^{-9}-10^{-10}$, which makes this
 scenario in principle accessible to the search of the $DEMON$ - dipole
 electric moment of neutron.
 \vspace{4mm}

  We are grateful to R.Barbieri, P.Fayet, H.Fritzsch, H.Leutwyler,
 A.Masiero, S.Pokorski, R.R\"{u}ckl,
 G.Senjanovi\'{c}, A.Smilga and K.Ter-Martirosyan
 for stimulating conversations. \\
 \pagebreak
 \newpage

  \end{document}